# Design and Market Considerations for Axial Flux Superconducting Electric Machine Design

Mark D. Ainslie[(1)], Ashley George[(2)], Robert Shaw[(2)], Lewis Dawson[(3)], Andy Winfield[(3)], Marina Steketee[(4)] and Simon Stockley[(5)]

(1) Bulk Superconductivity Group, Department of Engineering, University of Cambridge, Trumpington Street, Cambridge CB2 1PZ, UK

(2) Department of Materials Science & Metallurgy, University of Cambridge, Pembroke Street, Cambridge CB2 3QZ, UK

(3) Department of Physics, Cavendish Laboratory, University of Cambridge, J J Thomson Avenue, Cambridge CB3 0HE, UK

(4) Department of Chemical Engineering and Biotechnology, University of Cambridge, New Museums Site, Pembroke Street, Cambridge CB2 3RA, UK

(5) Judge Business School, University of Cambridge, Trumpington Street, Cambridge CB2 1AG, UK

Corresponding author e-mail address: mark.ainslie@eng.cam.ac.uk

*Abstract* – **In this paper, the authors investigate a number of design and market considerations for an axial flux superconducting electric machine design that uses high temperature superconductors. The authors firstly investigate the applicability of this type of machine as a generator in small- and medium-sized wind turbines, then the applicability as an in-wheel hub motor for electric vehicles. Next, the cost of YBCO-based superconducting (2G HTS) wire is analysed with respect to competing wire technologies and compared with current conventional material costs. Finally, different cooling options are assessed for the machine design.**

## I. INTRODUCTION

In 2010, the annual world electricity consumption was estimated at around 21,000 TWh, and this is estimated to increase to over 35,000 TWh by the year 2035 [1]. Given that there is a finite quantity of fossil fuel remaining, and the world's population continues to grow, our existing patterns of energy supply and usage are clearly unsustainable. In developed industrialised nations, the industrial sector uses about one third of energy consumed [2], and approximately two thirds of this energy is consumed by electric motors [3]. One way to reduce electrical energy consumption in electrical machines is the construction of electric motors and generators with better efficiency. Loss of electrical energy due to resistance to current flow, which is prevalent in conventional machines, translates directly to wasted energy and, therefore, to wasted economic resources.

Superconductivity offers zero to near zero resistance to the flow of electrical current when cooled below a particular cryogenic temperature. Consequently, the use of superconducting materials can improve the overall electrical system efficiency. In addition, superconducting materials are able to carry much larger current densities than conventional materials, such as copper. In electric machines, in particular, increasing the current and/or magnetic flux density increases the power density, which leads to reductions in both size and weight of the machine. The expected improved performance and efficiency, as well as smaller footprint, in comparison with conventional devices has seen continued interest in introducing superconducting materials to not only electric machines, but also to other electric power applications, such as transformers and cables.

Over many years of research, various superconducting machines have been shown to be technically feasible over a wide range of power ranges. For low temperature superconducting (LTS) materials, in particular, the complexity and cost of 4 K cryogenics hindered the commercial development of LTS machines [4], although there were a number of successful technical feasibility demonstrations [5]. The discovery of high temperature superconducting (HTS) materials in 1987 renewed enthusiasm for applied superconductivity research with the expectation that these materials could be exploited at 77 K, the boiling point of liquid nitrogen. Since then, a number of projects around the world have demonstrated the technical feasibility of HTS machines in various forms, including 5 MW and 36.5 MW motors for ship propulsion [6,7] by American Superconductor (now AMSC); a 380 kW motor [8], which was later developed into a 4 MW project machine [9], by Siemens; a 1.7 MW hydroelectric power generator by Converteam [10]; different HTS induction/synchronous motors at Kyoto University [11,12]; a 30 kW motor for an electric passenger car by Sumitomo Electric [13]; a 1 MW class synchronous motor for industry by KERI and Doosan Heavy Industries [14]; a 1 MW class podded ship propulsion motor by Kawasaki Heavy Industries [15]; and a sub-megawatt class propulsion system by Kitano Seiki [16].

However, both economic and technical challenges have meant that so far none of these machines have been commercialised. In this paper, the authors investigate a number of design and market considerations for an axial flux superconducting electric machine design that uses high temperature superconductors. This work was carried out as part of the University of Cambridge's Centre for Entrepreneurial Learning ETECH Project programme, designed to accelerate entrepreneurship and diffusion of innovations based on early stage and potentially disruptive technologies from the University. The axial flux motor design (for examples, see [17,18]), in general, provides higher torque/power density than other motor designs [18], and the use of superconductors is expected to improve these advantages even further [16]. In this study, the axial flux machine design is assumed to utilise high temperature superconductors in both wire (stator winding) and bulk (rotor field) forms, to operate over a temperature range of 65-77 K (liquid nitrogen temperatures), and to have a power output in the range from 10s of kW up to 1 MW (typical for axial flux machines), with approximately 2-3 T as the peak trapped field in the bulk superconductors.

Assuming that the main technical issues of this type of superconducting electric machine (chiefly, magnetising bulk superconductors in-situ and reducing AC loss in coils wound from HTS) can be resolved, there are still a number of issues that need to be considered for commercialising such technology, including:
1) Demand (closely related to application/s)
2) Price
3) Lifetime / reliability
4) Competition (other superconducting machine options, such as machines utilising a superconducting DC field winding)

5) Alternatives (conventional technology)
6) The state of the economy, in general

In order to investigate some of these issues, the paper is divided into four sections. In Section 2, the applicability of this type of machine as a generator in small- and medium-sized wind turbines and as an in-wheel hub motor for electric vehicles is investigated. In Section 3, the cost of YBCO-based superconducting (2G HTS) wire is analysed with respect to competing wire technologies and compared with current conventional material costs. In Section 4, different cooling options are assessed for the machine for the specified temperature range.

## II. APPLICATION STUDIES

In this section, the applicability of this type of machine as a generator in small- and medium-sized wind turbines is investigated, including the current and forecasted market and pricing for conventional turbines, as well as an in-wheel hub motor for electric vehicles.

### A. Small- & Medium-sized Wind Turbines

One potential application for this type of superconducting machine is as a generator in wind turbines, and the assumed power rating (up to 1 MW) puts this machine in the small- to medium-sized turbine category. The increased power density of a superconducting machine would mean a lighter generator, and for large machines in particular, there is a high demand for lighter generators, since at present the mass of the largest wind turbines exceeds the lifting capabilities of offshore installation vessels [19]. In addition, permanent magnet direct-drive (PMDD) wind turbines have received significant interest because elimination of the gearbox and slip rings can result in reduced down-time associated with maintenance and replacement, resulting in reduced cost-of-ownership and increased reliability. The low operating speed of a turbine generator require a high torque for a given power output, resulting in a physically larger machine [19]. However, the use of bulk HTS superconductors is expected to improve the machine performance even further, with trapped fields recorded as high as 17 T at 29 K [20] and up to 3 T at 77 K [21], allowing the removal of iron and an air-core design.

The wind turbine industry continues to grow and is rapidly expanding – the global wind turbine market saw 11% growth in capacity in 2010 with 39 GW delivered worldwide [22]. It is expected that by the end of 2016, global wind capacity will be close to 500 GW, in comparison with 237.7 GW at the end of 2012 [23]. Figure 1 shows the wind turbine market segmented by power output in terms of the expected amount of capacity installed each year from 2010 to 2025 [22]. In 2010, new turbines less than 1 MW accounted for 7% of the market (2.73 GW) and those greater than 3 MW accounted for 13%. In 2018, turbines greater than 3 MW are expected to account for 40% of the market, while turbines less than 1 MW are expected to shrink to 4% [22]. Although the wind turbine market on the whole is growing, new installations within the specified power range are expected to decrease, and the 1.5-2.49 MW class of wind turbine will dominate the market for the next five years [22].

The wind turbine market up to 1 MW is quite mature and there are a large number of companies producing conventional turbines of a high standard. Prices are dependent on the location of the site, but the following price ranges can be used as a guideline as a maximum for an equivalent superconducting design: £45k-65k for a 10 kW wind turbine (including installation) and £150k-250k for a 100 kW turbine (including installation).

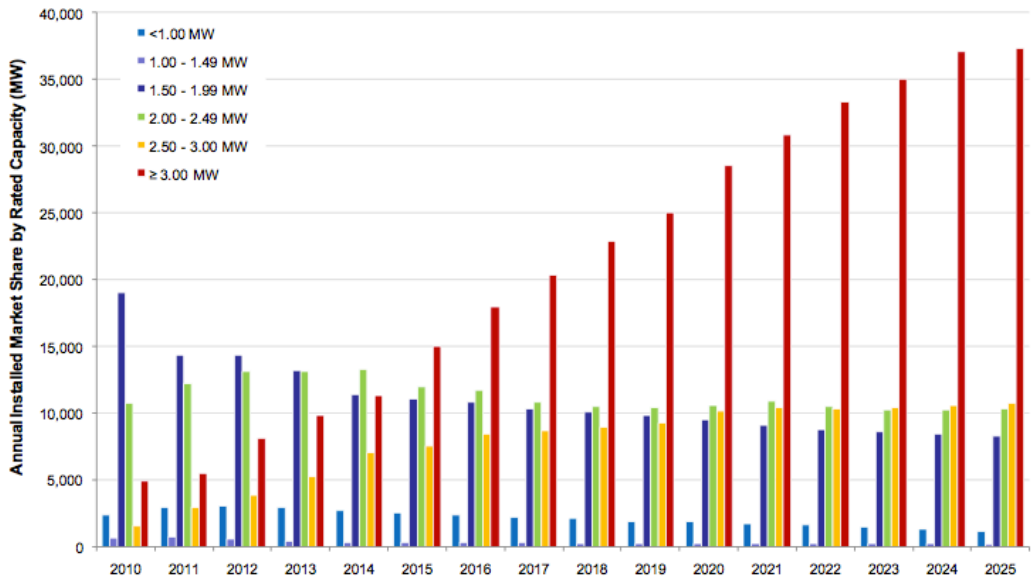

**Fig. 1.** Predicted global annual wind turbine installations by rated output power range, 2010-2025 [22].

*B. In-wheel Hub Motor for Electric Vehicles*

In-wheel hub motors are a concept gaining momentum for use in electric vehicles. This type of motor is commonly found on electric bicycles, and the idea is to place an electric motor into the unused space inside of a wheel to drive the wheel directly. There are a number of advantages: 1) manufacturers can remove the conventional engine bay, which allows for new and creative car designs, 2) removal of much of the powertrain (transmission, differential) results in a significant weight saving and reduced losses in and deterioration of mechanical transmission components, and 3) directly driving each wheel may improve car safety and dynamic drivability. In this section, we analyse the feasibility of an axial flux superconducting in-wheel hub motor for electric vehicles.

Figure 2 shows the electric vehicle global sales forecast for 2012-2017. For electric vehicles, a new component market is developing rapidly with growing electric motor sales [24]. The key feature here is how Original Equipment Manufacturers (OEMs, the automotive firms who manufacture components) are changing their 'make or buy' strategy for the electric vehicle market. Research suggests that they plan to directly purchase electric motors from external suppliers [25]. This presents an attractive opportunity for suppliers with sufficient manufacturing capabilities to enter the automotive value chain by supplying OEMs. This influences the business model a supplier should adopt and it is clear a number of competitors are reacting to this.

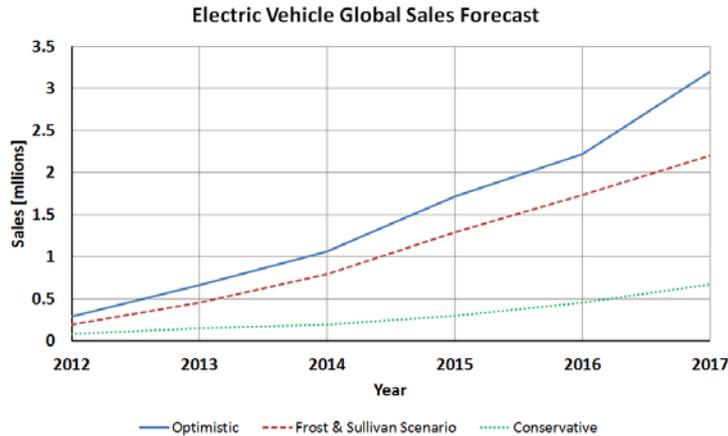
**Figure 2.** Electric vehicle global sales forecast for 2012-2017 [26].

There are several firms developing in-wheel hub motors at various stages of commercialisation and three companies are highlighted here. YASA Motors [27] is a spin-out from the Engineering Department at the University of Oxford, who are manufacturing high-torque axial flux motors. Typically spin-outs will go down a licensing route, essentially outsourcing their technical expertise since they lack the resources to manufacture themselves. However, YASA Motors have just received £1.45m of funding from a private investment firm whilst also securing their first contracts with major automotive OEMs [27]. This suggests they have adopted a business model to coincide with the 'make or buy' strategy of OEM's. Protean Electric is a US company manufacturing in-wheel motors, which is planning to have their latest prototype ready to present to OEMs by 2013, ready for volume production in 2014. They have just received US$84m in funding to proceed with building new manufacturing facilities in China [28]. This again highlights the business strategy of choice being to manufacture motors for distribution to OEMs, as well as the scale of the financial backing for rival technologies. Both of these motors do not utilise non-superconducting materials.

In terms of superconducting motors for electric vehicles, the closest to commercialisation is Sumitomo Electric, but this is not an in-wheel hub motor. Sumitomo first demonstrated the use of a superconducting motor to power a passenger vehicle in 2008 [13]. However, they have now shifted focus to application in buses, small trucks and forklifts, suggesting heavy duty usage may be a more profitable sector within the electric vehicle market for superconducting motors.

A business model can be identified corresponding to the changing value chain for electric vehicle components, i.e., raising capital from investors to fund manufacturing for sale directly to OEMs. However, many existing technologies already meet passenger vehicle requirements and any additional weight saved from the higher power density of a superconducting motor is negligible when compared to the weight saved through elimination of the powertrain. Furthermore, the likely cost, reliability and cooling issues mentioned in other sections means it is unlikely a superconducting motor would be feasible for this application. A suggested potential application would be heavy duty vehicles, such as trucks and plant machinery, and buses, which Sumitomo is targeting. The desire for higher power/torque densities for these functions means that cost and cooling system requirements become less of an issue. Although still many years away, electric aircraft may be an attractive application in the long term, since the power densities required can only be achieved with HTS motors [29].

# III. HTS WIRE COST

For a superconducting machine to be competitive with conventional machines, its cost-performance (C-P) ratio needs to be similar to or lower than that of conventional alternatives. In [30], for example, it is shown that although the superconducting machine (conventional armature winding, bulk YBCO rotor) provides improved performance with 3.5 times the torque density and the torque/mass ratio, the cost in comparison to a similar conventional machine, which is over an order of magnitude more expensive. A major component of the cost of a superconducting machine is the HTS material, and in large machines – for example, a 10 MW-class wind turbine generator – the cost of the system is dominated by the cost of the conductor, so a low cost conductor is critical [31]. Indeed, at current prices for HTS wire and cooling systems, the break-even point is around the 6-8 MW range for large scale wind turbines.

The C-P ratio for wire is commonly expressed in US dollars per kiloamp metre ($/kAm), which takes into account the raw wire cost (including labour and manufacturing) in addition to any improved performance under certain operating conditions. Competitive HTS wire requires a C-P ratio of approximately $20/kAm, based on the current C-P ratio of copper wire. 1G (BSCCO) and 2G (YBCO) HTS wire is compared with copper in Table 1.

Table 1. Summary of C-P ratio for HTS and conventional wires.

| Wire Type | C-P Ratio |
|---|---|
| Competitive Point | Approx. 20 |
| Copper | 15 – 55 |
| 1G (Di-BSCCO) | 180 – 230 |
| 2G (YBCO, current, SuperPower) | 450 [32] – 500 |
| 2G (YBCO, projected, SuperPower) | 175 [32] |

It is important to note that the superconductor C-P ratios above are for performance at 77 K in self-field (i.e., with no externally applied field). The maximum current a superconducting wire can carry (its critical current) is dependent on the magnetic field it 'sees'. Therefore, in the machine under analysis, both winding the wire into a stator coil (where the total field at any point in the coil is increased as the superimposition of the influence of the magnetic field at that point from the rest of the turns in the coil) and the trapped field in the bulk superconductors of 2-3 T would further reduce the performance of the superconducting wire and increase the C-P ratio. The in-field performance of BSCCO, which has a much lower irreversibility line, is inferior to that of YBCO, so the relative increase in its C-P ratio would be much greater.

The general consensus is that the current generation (2G) of HTS wire will not be competitive for some time, if at all, for the desired operating temperature range. However, this machine topology would be preferable cost-wise to a wound DC rotor topology (for example, [6,7]) because significantly less wire would be used in the overall design. The limiting factors still remain very much the same as those raised 15 years ago [33] and reasons include the inherent complexity of the manufacturing process, demand not meeting projections, and the rising raw material costs of metals such as copper (used in both BSCCO and YBCO) and silver (used in Di-BSCCO), both of which have seen their prices rise by about 500% over the past decade [34].

There are two main routes for making the cost of superconducting materials lower and more accessible: 1) materials improvements and 2) increasing/reducing production scale. 1) relates to the development of better materials that better exploit some physical phenomena, resulting in superconductors with higher critical temperatures, higher critical current densities, and the ability to carry higher currents in larger magnetic fields, and that are easy to fabricate. Although better devices can be designed on a case-by-case basis, superconducting devices will indirectly benefit from such material improvements. 2) relates to the scale of production for the materials, including both the scaling up of production, as well as the reduction of extraneous processes and equipment. Increased competition between existing wire manufacturers would also see prices reduced. In the case of HTS materials (both in bulk and tape forms), the question needs to be asked to what extent increases in the scale of production will reduce material costs, and indeed there are no studies to the authors' knowledge that investigate this thoroughly.

There is some promise for a next generation of round HTS wire – results have been reported recently on an isotropic, multifilamentary round wire Bi-2212 conductor that does not have extreme texture that (RE)BCO coated conductors possess, avoiding a high aspect ratio, large anisotropy and sensitivity to defects [35-37]. When a 100 bar overpressure is used to eliminate bubbles of residual gas in the conductor, the engineering critical current density, $J_e$, is the best of any available conductor above 17 T [35]. This round wire has significant potential to reduce the complexity and cost of building practical superconducting devices.

Another recent speculative development is the so-called 3G-HTS [38], which is produced from YBCO but in a novel manner such that the C/P ratio is expected to be in the range of $7-9/kAm [39]. The technology is based on ceramic silicone processing (CSP) to produce nanostructured composite HTS materials that could be as flexible as copper. Much time is likely to be needed to develop this technology, although it is unlikely to take as long as current processes took to develop, which have been refined since 1987. The company responsible for development of the aforementioned method, 3G-HTS Corporation [38], are currently developing a prototype manufactory, which if successful could represent a reasonable chance to reduce the cost of the superconducting machine. However, there is limited information available on the self- and in-field superconducting properties.

## IV. CRYOGENIC / COOLING OPTIONS

In order to maintain a superconductor's remarkable properties, a cryocooler or cryogenic system is required to keep the superconductor cold, at some operating temperature below its critical temperature, $T_c$. The discovery of high temperature superconductors in 1987 with $T_c$s greater than 77 K, the boiling point of liquid $N_2$, renewed interest in research on a number of different superconducting devices. However, while many superconducting devices have been shown to be technically feasible, one of the difficulties in commercialising such devices is the need for cryocoolers that are cheap with low weight/size and high power, efficiency and reliability. The importance of cryogenics as an enabling technology in the application of HTS technology has long been recognised [40]. Table 2 summarises current cryocooler options.

**Table 2.** Advantages and disadvantages of current cryocooler options.

| Cryocooler Type | Advantages | Disadvantages |
|---|---|---|
| Joule-Thomson | Steady flow | High pressure |
| | Vibration free | Low efficiency |
| | Large capacity | |
| Turbo-Brayton | Steady flow | Large heat exchanger required |
| | Vibration free | |
| | Large capacity | |
| Gifford-McMahon (GM) | Moderate cost | Large & heavy |
| | Low maintenance | Noise & vibrations |
| | Considerable manufacturing experience | |
| Stirling | Small size & weight | Prone to vibrations |
| | Moderate cost | No lubrication |
| | High efficiency | Long lifetime expensive to achieve |
| | High reliability (no moving parts) | |
| | Considerable manufacturing experience | |
| | Wide range of sizes available | |
| Pulse-Tube | Compact | Currently limited by efficiency, size & cost |
| | Highest efficiency between 40-200 K [29] | |
| | High reliability (no moving parts) | |
| | Reduced vibration | |
| | Wide range of cooling capacities available | |
| | Long-term promise (rapid advances being made) | |

The key issues in employment of cryogenic support technology have been identified as efficiency, reliability (usually defined by percentage of hours of downtime, which is affected by how much the machine vibrates and how many moving parts there are) and cost (per Watt of energy consumption) [40]. The latter three technologies in Table 2 have been identified as the most promising technologies in terms of the above three issues, and several examples of their use by existing companies in HTS technology application have been found:

- GM type: About 20,000 units per year are being produced [31]; Siemens used this type [41] in successful tests for a low-speed, high-torque HTS machine [42]; American Superconductor are looking into this option for their high capacity HTS wind turbines which are currently under development [43];
- Stirling type: Over 140,000 units have been manufactured to now [31]; Converteam chose this cryocooler type for use with their HTS hydroelectric generator [10];
- Pulse-tube type: This cryocooler type shows the most long-term promise, with rapid advances being made in terms of reliability and capacity [44]; Chubu Electric Power has developed such a cryocooler that can achieve 300 W at 77 K, with an efficiency of 6.7% and maintenance interval ten times that of earlier systems [45].

A final consideration is the choice of a cryogenic liquid. The preferred fluids are liquid helium (LHe), liquid neon (LNe) and liquid nitrogen ($LN_2$) because these are inert and do not combine with other fluids to create explosive mixtures [46]. Recent costs for each of these are as follows.

- $LN_2$    US$0.30-1.50 / litre [47]
- LHe    US$3.50-15 / litre [48]
- LNe    Can be more than 50 times the price of LHe

The evident choice is $LN_2$, based on cost and availability, and this is further supported by the existing infrastructure already in place for $LN_2$ supply and transport [40].

## V. CONCLUSIONS & RECOMMENDATIONS

In this paper, a number of design and market considerations have been investigated for an axial flux superconducting electric machine design that uses high temperature superconductors. The axial flux machine design is assumed to utilise high temperature superconductors in both wire (stator winding) and bulk (rotor field) forms, to operate over a temperature range of 65-77 K, and to have a power output in the range from 10s of kW up to 1 MW (typical for axial flux machines), with approximately 2-3 T as the peak trapped field in the bulk superconductors.

In terms of potential applications, although the wind turbine market on the whole is growing, new installations within the specified power range are expected to decrease to less than 4% by 2018. The likely cost and issues with reliability and cooling means it is unlikely a superconducting motor would be feasible for electric passenger vehicles, but a suggested potential application would be heavy duty vehicles, such as buses, trucks and plant machinery.

The general consensus is that the current generation (2G) of HTS wire will not be competitive for some time, if at all, for the desired operating temperature range. However, this machine topology would be preferable cost-wise to a wound DC rotor topology because significantly less wire would be used in the overall design.

Finally, it is suggested that the pulse-tube cryocooler is the most promising cryocooler option, leaving some time for further development, and liquid nitrogen is the most evident choice as a cryogenic liquid.

## ACKNOWLEDGEMENTS

Dr Mark Ainslie would like to acknowledge the support of a Royal Academy of Engineering Research Fellowship.